\documentclass[twocolumn, letter]{jpsj3}
\usepackage{txfonts}
\usepackage{graphicx}
\usepackage{dcolumn}
\usepackage{bm} 
\usepackage{color}
\definecolor{green}{rgb}{0.0, 0.5, 0.0}

\usepackage{ulem}

\def\vmu{\mbox{\boldmath $\mu$}}
\def\vecr{\mbox{\boldmath $r$}}

\def\runauthor{Y. Tomita and T. Kato}

\title{Relaxor behavior and morphotropic phase boundary in a simple model}

\author{Yusuke Tomita$^{1}$\thanks{ytomita@shibaura-it.ac.jp} and Takeo Kato$^{2}$}
\inst{$^1$ Shibaura Institute of Technology, 307 Fukasaku, Minuma-ku, Saitama-City,
Saitama, 337-8570\\
$^2$Institute for Solid State Physics, University of Tokyo, 5-1-5 Kashiwanoha, 
Kashiwa 277-8581}

\recdate{\today}

\abst{
A simple model to reproduce strong enhancement of dielectric response
near the morphotropic phase boundary (MPB) is proposed.
This model consists of long-range dipole-dipole interaction
and compositional chemical disorder incorporated by the variation in
lengths of dipole moments. By applying Monte Carlo simulation,
we show that there appears a ferroelectric boundary phase between 
two types of antiferroelectric phases at an optimal strength of randomness.
In the boundary phase, ferroelectric domain becomes remarkably large 
and flexible to external electric fields, leading to huge dielectric response.
This observation indicates that huge dielectric response near the MPB
originates from {\it local} polarization rotation under suppressed anisotropy 
by phase competition.
}
\kword{relaxors, dielectric response, domain structure, morphotropic phase boundary, 
Monte Carlo simulation}

\begin{document}
\maketitle

Ferroelectric relaxors based on perovskite oxides (ABO$_3$)
have attracted much interest because their characteristic dielectric properties
have several advantages to application.
The common feature of perovskite-type relaxors is intrinsic randomness due to 
compositional disorder in the arrangement of ions 
on the crystallographically equivalent sites.
For example, a typical ferroelectric relaxor 
Pb(Mg$_{1/3}$Nb$_{2/3}$)O$_3$ (PMN)~\cite{Smolenskii58} 
exhibits disordered arrangement of Mg$^{2+}$ and Nb$^{5+}$ at the B sites.
This intrinsic randomness is responsible for relaxor properties
such as huge permittivity with moderate temperature-dependence~\cite{Bokov06}.

In material survey of ferroelectric relaxors, it is well known that
structural phase transition induced by compositional change
plays a special role for obtaining excellent piezoelectric properties.
This phase boundary is known as the morphotropic phase boundary (MPB). 
For example, solid solution $(1-x)$Pb(Mg$_{1/3}$Nb$_{2/3}$)O$_3$-$x$PbTiO$_3$
(PMN-PT) has a rhombohedral(tetragonal) crystal structure for small(large) $x$, and 
shows large dielectric/piezoelectric response near MPB at $x\sim 0.35$~\cite{Park97}.
This feature is generally observed in relaxors based on solid solution of perovskite oxides;
the MPB is located for $(1-x)$Pb(Zn$_{1/3}$Nb$_{2/3}$)O$_3$-$x$PbTiO$_3$ (PZN-PT) 
at $x\sim 0.09$~\cite{Kuwata81}, and for PbZr$_{1-x}$ Ti$_x$O$_3$ (PZT) at $x \sim 0.5$~\cite{Jaffe71}. 

Study of the origin of large dielectric/piezoelectric response at the MPB 
is of much importance not only for understanding of relaxor properties,
but also for application to piezoelectric devices.
Huge dielectric/piezoelectric response at the MPB is intuitively understood
within the Landau-Ginzburg-Devonshire(LGD) theory~\cite{Devonshire54,Bell06,Ishibashi98,Yamada02}.
When different structural phases are almost degenerate near the MPB,
rotations of polarization become effective in response to external electric fields
instead of changing magnitude of polarization~\cite{Fu00,Vanderbilt01}. 
Recent discovery of a monoclinic phase near the 
MPB~\cite{Ohwada01,Ohwada03,Yokota09,Phelan10,Kaneshiro10} is consistent with this polarization
rotation scenario~\cite{Bell06,Fu00,Vanderbilt01}.
The LGD theory is, however, not satisfactory to describe the whole properties of relaxors
as it treats only spatially-averaged quantities.

In order to treat spatial dependence of polarization reflecting intrinsic atomic-scale randomness,
one of the most powerful methods is molecular dynamics (MD) simulation for an effective model
derived from first-principles calculation.
This approach has succeeded in quantitative evaluation of structure phase transition
in pure ferroelectric materials~\cite{Zhong94,Zhong95}, and has been also applied to relaxor 
ferroelectrics~\cite{Burton06,Tinte06,Burton08}. This approach is, however,
limited to short-time dynamics in the present computer resource, and is unable to 
access slow dynamics by large-scale domain motion, which is curtail to huge
dielectric/piezoelectric response at MPB. In the present situation, 
statistical-mechanical approach to a simplified model is still valuable to clarify 
what the essence of physics near MPB is.


In this paper, we propose a simple model, which can reproduce several
important properties near MPB~\cite{footnote1}. By applying
large-scale Monte Carlo simulation to this model, we calculate dielectric response,
domain structure, and structure factor. Our model exhibits intrinsic competition 
between different phases, which are connected by
polarization rotation. We show that the intrinsic atomic-scale disorder
strongly modifies the shape of domain structure at MPB (transition region between
phases). Our result indicates that in addition to the polarization
rotation mechanism, drastic change of domain shape at MPB is crucial
to huge dielectric/piezoelectric response. We also discuss its relevance 
to experimental results.

\begin{figure}[tb]
\begin{center}
\includegraphics[width=6.5cm]{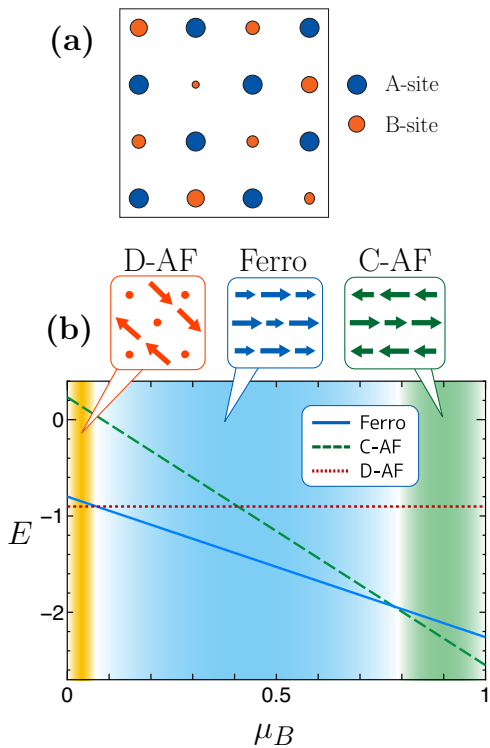}
\end{center}
\caption{(Color online)
(a) A schematic picture of an arrangement of A- and B-site moments.
Sizes of radii represent sizes of dipole moments.
(b) Plot of energies for D-AF (dotted line), Ferro (solid line),
and C-AF (broken line) configurations as functions of $\mu_{\rm B}/\mu_{\rm A}$, 
respectively.}
\label{fig:model}
\end{figure}

We constitute a simple model to describe MPB by the following strategy:
(1) The model is taken so that it exhibits competition between different phases
when magnitude of dipole moments is {\it uniform}, and then (2) atomic-scale
inhomogeneity is introduced by considering {\it random} distribution of magnitude 
of dipole moments. As a simplest model fulfilling these requirements,
we propose a model consisting of dipole moments 
on a two-dimensional square lattice:
\begin{equation}
{\cal H} = \sum_{i< j} \left[ \frac{\vmu_i \cdot \vmu_j}{r^3_{ij}}
- 3\frac{(\vmu_i \cdot \vecr_{ij})(\vmu_j \cdot \vecr_{ij})}{r^5_{ij}} \right]
\label{Hamiltonian}
\end{equation}
Here, $\vmu_i$ is a three-dimensional vector representing
an electric polarization caused by the ionic displacement at site $i$,
and $\vecr_{ij}$ is a displacement vector from site $i$ to $j$.
We further divide the square lattice into two sub-lattices, A and B as 
shown in Fig.~\ref{fig:model}~(a), and locate dipoles with different
magnitude on each lattice.

Let us first consider a {\it uniform} model, in which 
magnitudes of dipole moments are constant on each sub lattice.
We denote the magnitudes of moments on the A- and B-site
with $\mu_{\rm A}$ and $\mu_{\rm B}$, respectively. We show the
ground-state energies and configurations as a function of 
$\mu_{\rm B}/\mu_{\rm A}$ in Fig.~\ref{fig:model}~(b), assuming
that $2 \times 2$ periodicity like in Luttinger and
Tisza's preceding study~\cite{Luttinger46}.
As depicted in Fig.~\ref{fig:model}~(b), there are three ground states.
When the length of B-site moment $\mu_B/\mu_A$ is close to unity,
the ground state is a columnar-antiferroelectric (C-AF) state~\cite{Tomita09}.
On the other hand, the ground state is a diagonal-antiferroelectric (D-AF) 
as the ground state when $\mu_B$ is sufficiently small.
In the $\mu_B \to 0$ limit, the D-AF state can be regarded as another C-AF state on
the $\sqrt{2} \times \sqrt{2}$ square lattice constructed by A-sites.
It is noteworthy that directions of easy axis anisotropies
between the C-AF and the D-AF states differ from each other by 45-degrees.
That is, the length of B-site dipole moment affects directions of easy axes of dipoles on A-site.
Between the two AF states there is a wide ferroelectric region.
The ferroelectric order has an anisotropic easy axis except for $T=0$~\cite{Tomita09}.
The easy axis in this ferroelectric phase is vertical and horizontal near the C-AF
boundary, while it is diagonal near the D-AF boundary.

Analysis of the {\it uniform} dipole model has shown that
the present model exhibits different ordered phases controlled by
the ratio $\mu_{\rm B}/\mu_{\rm A}$.
We then introduce effect of atomic-scale randomness through spatial 
distribution of magnitude of dipole moments. Here, we
assume that magnitude of the A-site dipole moments is
fixed as unity ($\mu_{\rm A}=1$), whereas magnitude of the 
B-site dipole moments has spatial distribution. That is, we set 
magnitude of the B-site moment as
\begin{equation}
\mu = \left\{ \begin{array}{ll} 0.8 - 0.1n, & (n \le 7), \\
0 & (n \ge 8) \end{array} \right.
\end{equation}
where $n$ is an integer-valued random variable following
the Poisson distribution $P_{\lambda}(n)=\lambda^n e^{-\lambda}/n!$.
The strength of randomness is controlled by setting
the average value $\lambda$ of Poisson distribution.

Before showing analysis for the random dipole model,
we briefly mention relevance of the present model to the MPB observed
in perovskite oxides. In our model, the A-site dipoles corresponds to the
displacement of Pb ions, whereas the B-site dipoles to the displacement
of transition metal ions (the B-site in the perovskite structures). 
Since the intrinsic randomness is induced by compositional disorder
on the B-site, assumption of distribution in B-site dipole moments is not far from reality.
Then, $\lambda$ is related to strength of disorder, i.e., 
an compositional ratio $x$ between a relaxor ferroelectric material and 
a pure ferroelectric material. 
Our model has also been motivated from recent theoretical work 
focusing on importance of large polarization in the B-site~\cite{Grinberg07}.
We will discuss how the present effective model can be extended to 
more realistic one in the last part of this paper.

We performed Monte Carlo simulation in order to examine effects of the B-site randomness.
To realize an effective spin update in systems with long-range interactions,
O$(N)$ Monte Carlo method was adapted~\cite{Fukui09, Tomita09}.
The long-range dipole interactions were estimated by using the Ewald
summation method. We consider $32\times32$ square lattices throughout this paper.
To calculate dielectric susceptibilities, we apply an 
alternate external electric field with sinusoidal shape given as 
$E_{\rm amp}\sin(2\pi t/\tau)$ as a function of the Monte Carlo step $t$.
Though we cannot define {\it time} in Monte Carlo simulation,
one Monte Carlo step can approximately be regarded as forward time-evolution 
by a certain time step. In this paper, we set the magnitude and period of
electric field as $E_{\rm amp}=0.01$ and $\tau = 10^4$, respectively.
For thermalization, we discarded $10^5$ Monte Carlo steps while keeping on
applying external alternate field.
We executed $10^6$ Monte Carlo steps for measurement, so that
$10^6/\tau=10^2$ cycles are executed for measurement.
In order to average B-site randomness, 512 independent samples were simulated.

\begin{figure}[tb]
\begin{center}
\includegraphics[width=6.5cm]{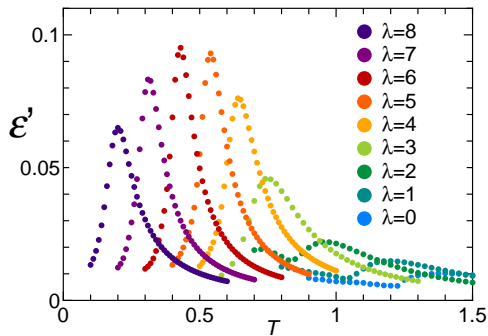}
\end{center}
\caption{(Color online)
Plot of real part of dielectric susceptibilities per site for several $\lambda$'s.
Data of each $\lambda$'s are horizontally shifted to improve the visibility.
Real temperature $T_{\rm R}$ for each $\lambda$'s are given by
$T_{\rm R} = T - 0.1(8 - \lambda)$, where $T$ represents a temperature 
given in the figure.}
\label{fig:epsilon}
\end{figure}

\begin{figure*}[tb]
\begin{center}
\includegraphics[width=16cm]{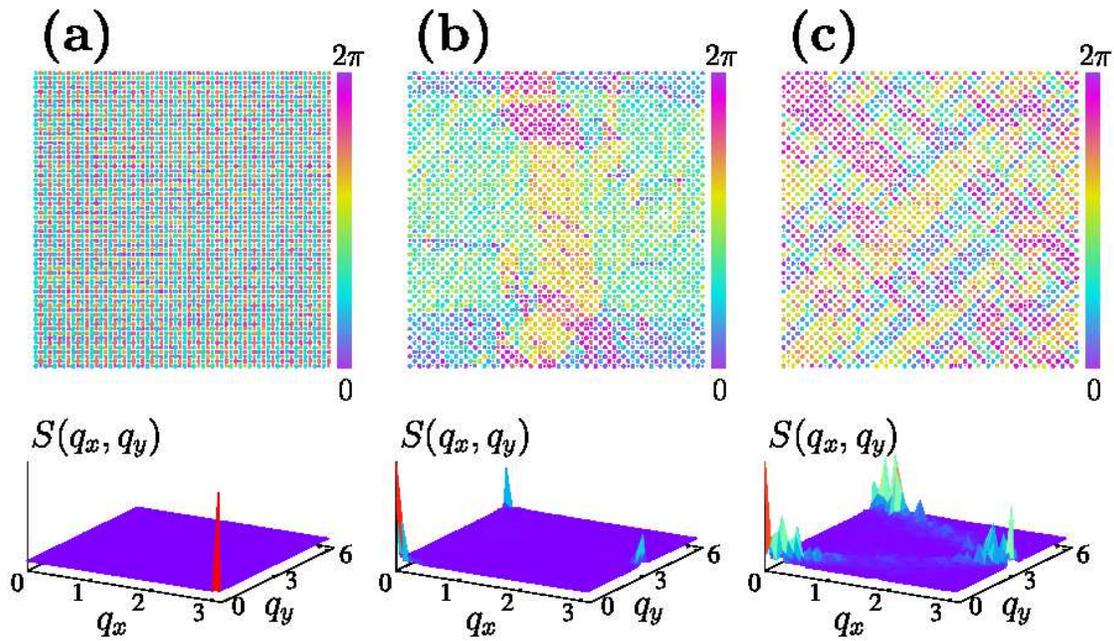}
\end{center}
\caption{%
(Color online) Snapshots of dipole configurations projected on to the lattice 
plane (upper panels) and structure factors (lower panels)
for (a) $\lambda = 0$, (b) $\lambda = 5$, and (c) $\lambda = 8$ at $T = 0.1$.
Sizes of radii represent sizes of projected dipole moments, and
colors of circles indicate directions of dipole moments on the lattice plane.
Almost dipole moments reside in the lattice plane except for those
located adjacent to vortices. }
\label{fig:domain}
\end{figure*}

In Fig.~\ref{fig:epsilon}, we shows real part of calculated dielectric susceptibilities 
per site for several values of $\lambda$ as a function of the temperature $T$.
As seen from the figure, the dielectric susceptibilities has a maximum at $T = T_{\rm max}$,
and the maximum temperature $T_{\rm max}$ is lowered as $\lambda$ increases.
The peak of the dielectric susceptibility is well suppressed for small $\lambda$, 
whereas it rapidly grows with increasing $\lambda$. The maximum value 
becomes largest at an optimum value $\lambda\sim 5-6$, and it is reduced
for larger value of $\lambda$. These features well resemble
the behavior of the dielectric response near MPB in perovskite oxides,
when we think $\lambda\sim 5-6$ to be the MPB in the present model.

To examine the mechanism of emerging of the maximum,
snapshots and structure factors $S(q_x,q_y)$ for $\lambda = 0$, $5$, and $8$
are shown in the upper and lower panels of Fig.~\ref{fig:domain}, respectively. 
For $\lambda = 0$, the columnar-antiferroelectric (C-AF) phase is stable at low 
temperatures (Fig.~\ref{fig:domain}~(a)). 
The long-range C-AF ordering is indicated by a sharp peak at $(q_x,q_y) = (\pi,0)$,
and most of the width of the ferroelectric domains are a unit lattice length
as seen from the snapshot, explaining suppression of dielectric susceptibilities for small $\lambda$.
On the other hand, a relaxor ferroelectric (R-F) phase is realized
at $\lambda = 8$ at low temperatures (Fig.~\ref{fig:domain}~(c)).
One broad peak of the structure factor at $(q_x,q_y) = (0,0)$ indicate ferroelectric short-range ordering,
the other peak at $(q_x,q_y) = (\pi,\pi)$ reflects the difference of magnitude of moments
on the A- and B-sublattice. Here, we point out that at finite temperatures,
direction of polarization in ferroelectric domain is affected by the neighboring phases.
At $\lambda = 8$, polarization in ferroelectric domain is restricted along the diagonal
direction affected by the D-AF phase. In this phase,
strong randomness make peaks in the structure factor diffusive, i.e., extended 
along the diagonal direction, leading to butterfly-shaped structure factors which are 
typically observed in relaxors. These properties characteristic of the R-F phase
explains relatively large dielectric response for $\lambda > 6$ in our model.

Here, we note important difference between the C-AF phase and R-F phase. 
The former phase realized for small $\lambda$ has
easy axis of dipole moments along $(1,0)$ or $(0,1)$ direction, whereas the latter 
phase realized for large $\lambda$ along the diagonal direction $(1,1)$ and $(1,-1)$. 
This difference can be seen by the shape of domains shown in the 
upper panels in Fig.~\ref{fig:domain}~(a) and (c). We stress that transition between
these two phases necessarily accompanies {\it polarization rotation}~\cite{Fu00}. 
What happens if $\lambda$ is set on the boundary between the above two phases?
This is the main problem which we will clarify from now.

The boundary between the C-AF and R-F phases is located 
around $\lambda = 5$ (Fig.~\ref{fig:domain}~(b)). 
At $\lambda=5$, ferroelectric ordering remains in this boundary region 
as indicated by two peaks in the structure factor. 
The ferroelectric domain size becomes, however, much larger as seen from the snapshot. 
The remarkable enhancement of correlation length 
for ferroelectric ordering is also indicated from the sharpness of the peak in the structure 
factor at $(q_x,q_y) = (0,0)$; the correlation length
becomes much larger than the case of $\lambda = 8$, for which diffusive 
butterfly-shaped pattern is induced by local atomic-scale domain.
A noteworthy fact is that directions of the polarization in ferroelectric
domains are not restricted in contrast to the C-AF and R-F phase.
This feature can be seen also from the shape of the domain walls, which
are irregular shaped and much rounder than in the C-AF and R-F phase. 
The free-wheeling of the polarizing directions in the intermediate region
comes from competitions between the C-AF and the R-F states.
Because of the B-site randomness, the competitions differ from area to area.
In view of the local stability, a favorable dipole direction
is affected by neighboring B-site moments.
However, the direction of the ferroelectric domain
that the dipole belongs to also affects the favorable direction.
Therefore, favorable directions of dipoles can be changed easily
by a small disturbance, so that dipoles react immediately to
the external electric field in the intermediate phase. This competition is the reason 
why the largest susceptibility peak is achieved in the intermediate region.

\begin{figure}[tb]
\begin{center}
\includegraphics[width=6cm]{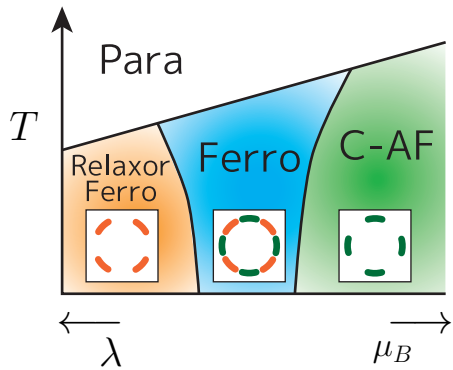}
\end{center}
\caption{(Color online) A schematic phase diagram of our model.
At sufficiently high temperatures, there is a paraelectric (Para) phase.
Phases at low temperatures depend on
the average length of B-site moments $\mu_{\rm B}$.
Arcs in three squares represent favorable directions of dipoles.}
\label{fig:phasediagram}
\end{figure}

In Fig.~\ref{fig:phasediagram} we summarize a schematic phase 
diagram of our model. Since an increase in size of dipole moment raises
transition temperature, phase boundary adjacent to the paraelectric phase is 
proportional to the average size of $\mu_{\rm B}$.
Stable dipole configurations are influenced by $\mu_{\rm B}$
(see Fig.~\ref{fig:model}(b)), so that we observed the R-F, ferroelectric,
and C-AF phases as increasing $\mu_{\rm B}$ (decreasing $\lambda$).
Due to the configurational entropy, ferroelectric phase becomes wider
as increasing temperature. In the present model, 
the B-site randomness controls not only the stability of the C-AF and R-F
phase, but also changes the easy axis of dipole moments.
A schematic figure of the easy-axis potential on dipole moments is also 
shown in Fig.~\ref{fig:phasediagram}. In the intermediate region,
the easy-axis potential is mixture of the ones of the C-AF and R-F phase.
The resulting mixed potential forms the dimple at the bottom of a wine bottle,
and it makes dipoles easy to rotate.

Our result is consistent with polarization rotation mechanism proposed by
Fu and Cohen~\cite{Fu00}, which claims multiple ordering energetically 
degenerate near MPB. Their discussion is, however, based on the spatially
averaged quantities as well as other theoretical works based on the
LGD theory, and couldn't predict anything on domain structures.
Indeed, our Monte Carlo simulation has demonstrated that change 
of anisotropy in dipole moments does strongly affects spatial structure of
ferroelectric domains. This result indicates that {\it local} polarization rotation 
under suppressed anisotropy due to phase competition makes domain wall
{\it flexible} to external field, leading to huge dielectric response.
This is the main result of this paper.

Our observation of large ferroelectric domain formation 
near the MPB is expected to be examined by experiments
such as neutron scattering~\cite{Matsuura06,Ohwada11,Matsuura11} and 
transmission electron microscopy (TEM)~\cite{Fu09,Bokov04,Ye99,Ye00,Kurushima11}.
Recently, Matsuura et al. observed that correlation length becomes large
near the MPB~\cite{Matsuura06}. The same group further fortified
the mechanism by inelastic neutron scattering measurements
and an analysis of a pseudospin-phonon coupled model~\cite{Matsuura11},
which is a similar scenario of our model. 
An anomalous nature of ferroelectric domain at the MPB has also been reported
in the TEM experiment~\cite{Kurushima11}.

Finally, we discuss extension of our simple model toward more realistic models.
It is straightforward to extend our model to three-dimensional models without
any problems except for computational resources. In our model, we omitted
short-range interaction related to covalent bonding between neighboring ions
and strain of the surrounding media. As a result,  the C-AF phase has been 
realized in our model in the pure material side (small-$\lambda$ region), instead 
of pure ferroelectric phase usually realized in perovskite oxides. This artifact can be 
corrected by considering short-range interaction, though it is left for future work.
We, however, claim that essence of physics at MPB should come
from {\it local} polarization rotation due to suppressed anisotropy by mixture of 
anisotropy potentials, which are sufficiently taken into account in the present model.

In summary, we proposed a simple dipole model, and executed
Monte Carlo simulations to it. We showed that there appear a boundary 
ferroelectric phase between the C-AF and R-F phases at an optimal B-site 
randomness, and that it has large ferroelectric domains with flexible
walls. This novel feature is due to mixture of two different easy axes 
neighboring phases. We expect that our model explains the reason 
why the realization of the large dielectric response around MPB regions
of a wide variety of binary and ternary Pb-based solid solutions.

We thank Kenji Ohwada, Masato Matsuura, and Shigeo Mori for valuable
discussions.
The computation in the present work is executed on computers at the
Supercomputer Center, Institute for Solid State Physics, University of Tokyo. 
The present work is financially supported by MEXT Grant-in-Aid for
Scientific Research on Priority Areas
``Novel States of Matter Induced by Frustration'' (19052002).


\begin{thebibliography}{99}
\bibitem{footnote1}
A similar approach has been employed by the authors to explain a phase diagram and 
dielectric response of Pb(In$_{1/2}$Nb$_{1/2}$)O$_3$ (PIN)~\cite{Tomita10,Ohwada10}.
\end{thebibliography}
\end{document}